\documentclass[aps,prl,reprint]{revtex4-1}
\usepackage{graphicx}
\usepackage{epstopdf}

\begin{document}

\title{Homes Scaling in Ionic Liquid Gated La$_{2}$CuO$_{4+x}$ Thin Films}

\author{J. Kinney}
\author{J. Garcia-Barriocanal}
\author{A. M. Goldman}

\affiliation{School of Physics and Astronomy, University of Minnesota, 116 Church
Street SE, Minneapolis, Minnesota 55455, USA}

\date{\today}

\begin{abstract}
Finding more efficient ways of exploring the doping phase diagrams
of high temperature superconductors as well as probing the fundamental
properties of these materials are essential ingredients for driving
the discovery of new materials. We use a doping technique involving
gating with ionic liquids to systematically and continuously tune
the $T_{c}$ of superconducting La$_{2}$CuO$_{4+x}$ thin films.
We probe both the transport properties and the penetration depth of
these samples and find that Homes scaling $\lambda^{-2}\propto\sigma T_{c}$
is obeyed, consistent with these materials being in the dirty limit.
This result is independent of the precise mechanism for the gating
process as all of the parameters of the scaling relationship are determined
by direct measurements on the films.
\end{abstract}


\maketitle
The study of high-temperature superconductivity is a field dating
back to the discovery of LaBa$_{2}$Cu$_{3}$O$_{7-x}$\cite{Bednorz1986}.
Since that time, there has been a steady drive to increase the transition
temperature of these materials with the ultimate goal of finding a
room temperature superconductor which could unlock many practical
technologies. At a basic level this push for higher transition temperatures
is a material search. Ionic liquids offer the exciting possibility
of increasing the efficiency of this search, allowing one to explore
a whole range of doping for a material system in a single sample,
where previously a whole series of sample growths would be required.
We demonstrate the use of ionic liquids in an electric double layer
transistor configuration to continuously tune the transition temperature
of superconducting La$_{2}$CuO$_{4+x}$ (LCO) thin films. This allows
us to precisely monitor how the change in the carrier concentration
affects the transition temperature, resistance and superconducting
penetration depth of the material. We highlight these details and
demonstrate through the verification of Homes scaling\cite{Homes2004}
that that this approach is nevertheless a powerful alternative to
chemical doping for systematically studying the properties of high-temperature
superconductors. This is true despite the fact that the doping in
this material may not be a purely electrostatic effect. 

These films were grown by ozone-assisted molecular beam epitaxy \cite{Berkley1988}.
We use Cu and La Knudsen effusion cells to produce stable rates for
these two materials. The rate for each source is measured prior to
the film growth using a quartz crystal microbalance. We alternately
deposit one half of a unit cell of La then Cu by opening shutters
for each source. The film presented here is 3 unit cells thick. The
substrate temperature is 700 $^{0}C$ as measured by a Williamson
Pro 92-38 pyrometer and under an ozone pressure of 3x10$^{-5}$ torr
during growth. We monitor the growth of the films using RHEED and
see strong 2D film streaks epitaxially locked to the\textit{ a} and
\textit{b} axes of the substrate's lattice. We use $<001>$ oriented
SrLaAlO$_{4}$ substrates supplied by MTI corp. The $T_{c}$ of the
initial sample before the application of ionic liquid and gate voltage
is well-defined and is greater than 40K, which is a sign of high quality
and implies that the excess oxygen is near its maximal value of x=0.12.

Tuning the properties of high $T_{c}$ materials using a conventional
FET configuration is a well established technique for systematically
adjusting the carrier concentration of superconducting materials\cite{Matthey2007,Salluzzo2008,Rufenacht2006b}.
The use of ionic liquids to tune the properties of thin films has
grown rapidly in the past several years due to the possibility of
producing large changes in the carrier concentration of a material.
Some of the first efforts in this regard were by Iwasa's group \cite{Ueno2008}.
In terms of studying high temperature superconductors, this technique
has been applied to La$_{2-x}$Sr$_{x}$CuO$_{4}$\cite{Bollinger2011},
YBa$_{2}$Cu$_{3}$O$_{7-x}$\cite{Leng2011,Leng2012}, LCO\cite{Garcia-Barriocanal2013}
and Pr$_{2-x}$Ce$_{x}$CuO$_{4}$ \cite{Zeng2015}. In all of these
studies the focus was on the superconductor-insulator transition.
The work presented here is an extension of that done on LCO with the
technique of penetration depth measurements added in parallel with
standard transport measurements. This combination allows us to determine
the evolution of the relationships between various superconducting
and normal state properties with applied gate voltage and temperature.

\begin{figure}
\includegraphics[width=1\linewidth]{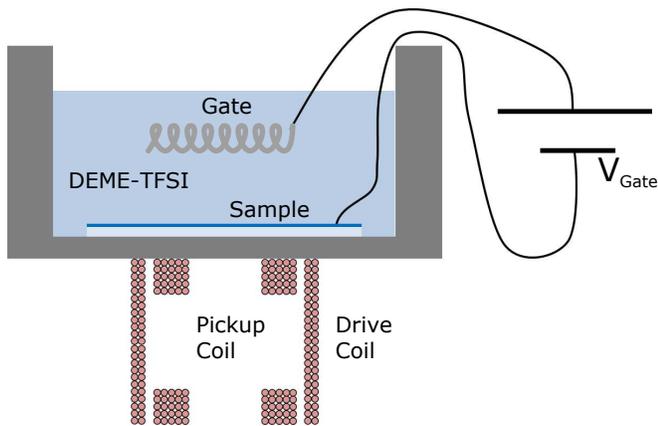} \protect\caption{(color online) Schematic of the sample setup. The entire sample is
immersed in the ionic liquid DEME-TFSI. A gate electrode made of Pt
is also immersed in the ionic liquid. Not pictured are leads going
to the corners of the sample for making resistance measurements. The
sample holder is made of plastic (Delrin) to minimize extraneous screening
currents. The two-coil assembly is composed of an outer drive coil
and and inner pickup coil. The inner pickup coil consists of two oppositely
wound sections to minimize the direct mutual inductance between the
drive coil and the pickup coil. The sample is patterned into a disk
with a 5 mm diameter. The inner diameters of the pickup coil and drive
coil are 2 mm and 3 mm respectivlty. The gate is a coil of Pt wire
with a surface area double that of the sample. \label{SampleSetup_fig}}
\end{figure}

We used a two-coil mutual inductance technique for tracking the penetration
depth \cite{Hebard1980,Jeanneret1989,Lin1995a,Turneaure1996,Turneaure1998}.
The experimental setup is shown in Fig. \ref{SampleSetup_fig}. Both
the drive and pickup coils are located on the back side of the sample.
The holder is constructed out of Delrin. We use an SRS SR830 lock-in
amplifier to both produce the drive signal and in conjunction with
an SRS SR560 amplifier to measure the output from the pickup coil.
The AC drive signal has a frequency of 50 kHz and an amplitude of
1 mA. We use a numerical procedure for determining the penetration
depth of the sample from the measured real and imaginary components
of the voltage read by the pickup coil. We model the film with a complex
impedance, $Z=R+i\omega L_{k}$, where R and $L_{k}$ are the sheet
resistance and the kinetic inductance of the film, respectively. In
the case of a thin film with thickness d and $\lambda\gg d$, the
kinetic inductance is directly related to the London penetration depth
as $L_{k}=\mu_{0}\lambda^{2}/d$, where $\mu_{0}$ is the permeability
of free space and $\lambda$ is the London penetration depth.

\begin{figure}
\includegraphics[width=1\linewidth]{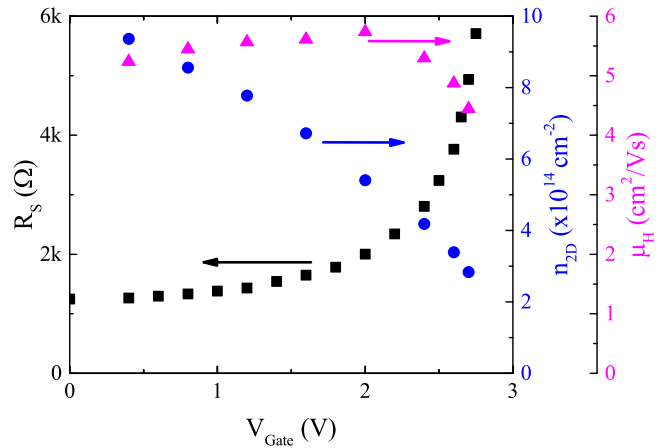} \protect\caption{(color online) Plot showing how $R_{S}$(squares), $n_{2D}$(circles),
and $\mu_{H}$(triangles) evolve as the gate voltage is increased.
All three quantities were measured at 180 K. We see that as we increase
the gate voltage the resistance steadily increases and the number
of free carriers decreases. We also note that, while the mobility
does begin to decrease above 2.5 V, we are primarily affecting the
carrier concentration of the film and for the most part the mobility
is remaining constant.\label{gating_fig}}
\end{figure}

We used the ionic liquid DEME-TFSI to tune the properties of our superconducting
samples. The ionic liquid is initially applied to the surface of the
sample at room temperature but all further gating is done at 245 K.
Figure \ref{gating_fig} shows how the resistance, carrier concentration,
and mobility evolve with the applied gate voltage at a temperature
of 180K. All three quantities were determined using a 4-wire resistance
technique in a Van der Pauw configuration. Each measurement step involves
warming the sample and ionic liquid to 245 K, which is above the melting
point of the liquid. The gate voltage is then changed at this elevated
temperature and allowed to relax for 20 min. We apply a positive voltage
to the gate, which corresponds to removing holes or reducing the number
of charge carriers in the sample. As the gate voltage is increased
we see both an increase in the resistance of the film and a decrease
in the carrier concentration. While the mobility does begin to decrease
at higher gate voltages we see that the primary effect of the applied
voltage is to reduce the number of free holes in the material. We
note that the gating procedure is not purely electrostatic as the
changes upon gating are irreversible. The evidence for this conclusion
comes from removing the applied gate voltage and observing that the
resistance does not return to the zero-voltage state. In fact, we
must apply a negative gate voltage in order to return the resistance
to its initial state. There is evidence in many other material systems
doped with ionic liquids showing that the gating process is often
a mixture of electrochemical and electrostatic effects \cite{Jeong2013,Jeong2015,Lang2014,Petach2014}.
Particular to LCO, there is evidence that the interstitial oxygens
are mobile down to 200 K\cite{Fratini2010} and it would not be surprising
to think that the ionic liquid gating procedure removing holes actually
removed some oxygen. Now, while there are advantages to a purely electrostatic
process, namely reversibility and no introduction of disorder, an
\textit{in-situ} electrochemical doping method, which we believe may
be involved in the process presented here, is still a powerful and
useful tool.

\begin{figure*}
\includegraphics[width=1\linewidth]{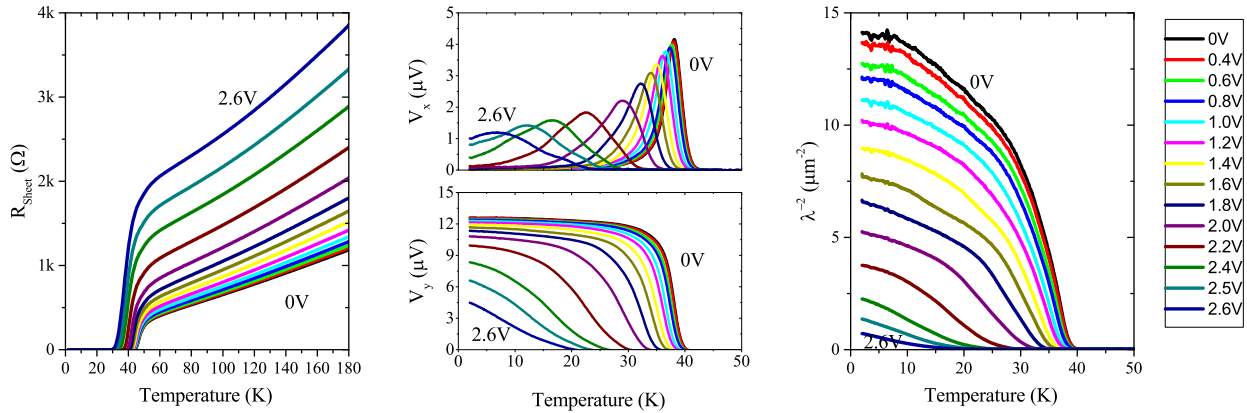} \protect\caption{(color online) (Left) The sheet resistance of a 3 unit cell LCO sample
tuned using ionic liquids. The initial, 0V, state has the highest
$T_{c}$ and the lowest resistance. As the gate voltage is increased
$T_{c}$ decreases and the normal state resistance increases. (Center)
The real and imaginary components of the voltage in the pickup coil.
The real component, $V_{x}$, is peaked at $T_{c}$ of the film and
becomes broader with increasing gate voltage. The imaginary component,
$V_{y}$, is a half down which steadily grows below $T_{c}$. (Right)
The inverse of the London penetration depth squared, which is proportional
to the superconducting electron density, grows with decreasing temperature
below $T_{c}$. \label{sample_RVL_fig}}
\end{figure*}

Due to imprecise knowledge of the distance between the coils and the
sample, we cannot determine the absolute size of the penetration depth.
Instead, to set the scale, we use 2700 \AA{} based on literature values
for La$_{2-x}$Sr$_{x}$CuO$_{4}$\cite{Startseva1998,Aeppli1987,Lemberger2011}
a compound with a similar transition temperature, to calibrate the
unknown spacing between the sample and the coils. We then measure
changes in the penetration depth from that starting point as we adjust
the doping of the sample using the gate electrode in the ionic liquid.
The real and imaginary components of the voltage in the pickup coil
are shown in Fig. \ref{sample_RVL_fig}. The real component has a
peak centered at the superconducting transition and the width of the
peak is, in part, a measure of the homogeneity of the sample at that
doping. We see that with increasing applied gate voltage the peak
moves to lower temperatures and broadens. We interpret this as a change
in the doping of the material, removing holes, but not necessarily
in a fully homogeneous way. We will use the center of this peak as
our definition of $T_{c}$ for the purposes of checking the Homes
scaling relation \cite{Homes2004}. The imaginary component of the
voltage, while not explicitly equal to it, represents the kinetic
inductance of the sample which is directly related to both the penetrations
depth and superconducting electron density. We see the imaginary component
move to lower temperatures and go to zero with increasing gate voltage.
Data on the resistance of the sample can be obtained in parallel with
the two-coil signal. We see that as we increase the gate voltage and
remove holes the normal state resistance increases and $T_{c}$ steadily
decreases. We can translate the combination of real and imaginary
components of the pickup coil signal into the penetration depth of
the sample.

The first thing to check is the temperature dependence of the penetration
depth. Depending on the pairing state of the superconductor we expect
$\Delta\lambda$ to have a specific temperature dependence. For an
\textit{s-}wave pairing state with a full gap we expect an exponential
dependence for $\Delta\lambda$. For a \textit{d-}wave pairing state,
where there are nodes in the gap, we expect a linear dependence for
$\Delta\lambda$. In the case of strong scattering this linear dependence
shifts to $\Delta\lambda\propto T^{2}$ \cite{Prohammer1991}. We
have seen (not shown) that this sample exhibits a quadratic temperature
dependence, which implies scattering is playing a significant role.
The ultra thin nature of this sample means that the upper and lower
surface roughness make up a large fraction of the thickness of the
entire film. We used both X-ray reflectivity (XRR) and atomic force
microscopy (AFM) to characterize the film roughness. From XRR our
estimates of the substrate-sample interface roughness and the sample
surface roughness are 3 \AA{} and 8 \AA , respectively. Using AFM
we measure the sample surface roughness to be 3 \AA . All of these
quantities are significant compared to the c-axis lattice parameter,
which is 13.3 \AA . Considering that the entire film thickness is
3 unit cells and the roughness of both of the top and bottom surface
are significant fractions of a unit cell it is reasonable to expect
this film to be in the strong scattering limit.

\begin{figure}
\includegraphics[width=1\linewidth]{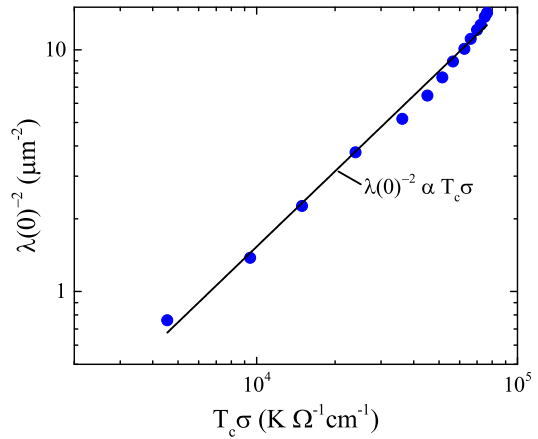} \protect\caption{(color online) $\lambda(0)^{-2}$vs $T_{c}\sigma$ for a 3 unit cell
LCO film. The properties of the film were systematically tuned using
the ionic liquid and gate electrode setup pictured in Fig. \ref{SampleSetup_fig}.
We see a linear relationship between $\lambda(0)^{-2}$and $T_{c}\sigma$
in agreement with the scaling relationship seen by Homes \textit{et
al}. \label{Homes_fig}}
\end{figure}

The next check we can make with this data set is to see if the Homes
scaling relationship is obeyed\cite{Homes2004,Homes2005}. Figure
\ref{Homes_fig} demonstrates that $\lambda^{-2}\propto T_{c}\sigma$
does in fact provide a good description of our data. We use the conductivity
at 180 K and define $T_{c}$ using the peak value of the real component
of the two-coil signal. Both of these are clearly defined values even
when the transition becomes broad at higher gate voltages. The scaling
relationship also holds when the conductivity just at the onset of
superconductivity is employed. Following the work of Zuev \textit{et
al.}\cite{Zuev2005} we can use the relationship 
\begin{equation}
k_{B}T_{2D}=\frac{\Phi_{0}^{2}}{8\pi\mu_{0}}\frac{d}{\lambda^{2}(T_{2D})},\label{eq:BKT}
\end{equation}
 to define $T_{2D}$, which is the Berezinskii-Kosterlitz-Thouless-transition
temperature. Here $\Phi_{0}$ is the flux quntum. If we use the entire
film thickness of 40 \AA{} as the value of $d$ we find good agreement
between this definition of $T_{2D}$ and the value of $T_{c}$ we
determine from the peak of the real component of the two coil signal. 

There are a number of different proposals for the interpretation of
the $\lambda^{-2}\propto\sigma T_{c}$ relationship \cite{Imry2012,Kogan2013,Taylor2007,Basov2011}.
The simplest argument comes from the fact that the film is in the
strong scattering limit ($\Delta\ll h\tau^{-1}$) to begin with. For
our films we use a simple Drude model to estimate the scattering rate.
Taking the values of $R_{s}$ and $n_{2D}$ from Fig. \ref{gating_fig}
and the relationship $\frac{1}{R_{s}}=\frac{n_{2D}e^{2}\tau}{m}$
we can calculate $h\tau^{-1}$ to be 1.3 eV, which is much larger
than the superconducting gap in these types of materials \cite{Hashimoto2014}.
It is important to note that while the Homes data set is a compilation
of many samples measured over the course of many years, the data shown
here is from a single sample continuously doped to cover a range of
$T_{c}$ values in a systematic and efficient manner. This points
to the power of using ionic liquids in an electric double layer transistor
configuration to systematically study the fundamental properties of
superconducting samples. 

\begin{figure}
\includegraphics[width=1\linewidth]{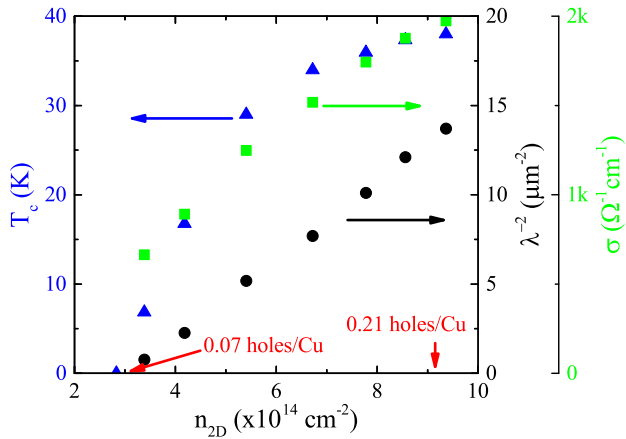} \protect\caption{(color online) Plot showing how $T_{c}$(triangles), $\lambda^{-2}$(circles),
and $\sigma$(squares) evolve as a function of doping in a 3 unit
cell LCO thin film. For reference we include markers at the top and
the underdoped edge of the superconducting dome, where the conversion
from charges per square cm to holes per copper is done using the fact
that we have a 3 unit cell sample which corresponds to 6 copper oxygen
planes.\label{doping_fig}}
\end{figure}

The initial motivation for exploring whether there is a fundamental
relationship between $T_{c}$ and $\lambda^{-2}$ was derived from
the empirical relationship proposed by Uemura \textit{et al}. \cite{Uemura1989}.
The data presented here does not follow the relationship proposed
by Uemura of $T_{c}{}^{2}\propto\lambda^{-2}$. This is actually readily
seen by a more careful examination of Fig. \ref{doping_fig}. We see
that the quantity $\lambda^{-2}$, which is proportional to the superconducting
electron density, increases linearly with the number of free carriers
in the normal state. Simply, the number of superconducting electrons
in the superconducting state is proportional to the number of free
holes in the normal state. This means that we could simply switch
the x-axis of Fig. \ref{doping_fig} from doping (carrier concentration)
to $\lambda^{-2}$ without effecting the shape of the $T_{c}$curve.
This means the relationship between $T_{c}$ and $\lambda^{-2}$ will
just be a remapping of the superconducting dome and therefore not
a linear function as seen by Uemura in other, chemically doped high-$T_{c}$
materials. There is precedent for a sub-linear relationship for $T_{c}\propto[\lambda^{-2}]^{x}$,
where a value of $x=0.5$ follows from an argument about critical
scaling where there is a quantum critical point (QCP) at the edge
of the dome\cite{Zuev2005,Lemberger2011}. While we do see a sub-linear
relationship ($x<1$), it is not a pure single power law over the
range of dopings examined here, much of which is likely far from the
QCP point. 

In summary, we have used the ionic liquid DEME-TFSI to systematically
tune the transition temperature of under-doped LCO thin films, and
have displayed and analyzed data from a 3-unit cell thick film. This
has allowed us to carefully track how the resistance, Hall Effect,
and penetration depth are interrelated. Our results are in good agreement
with the Homes scaling relation consistent with the fact that this
film is in the strong scattering limit. In addition, this unique setup
allows us to explicitly show that the super-electron density, as measured
by $\lambda^{-2},$ is linearly proportional to the free carriers
in the normal state. As a result of this linear dependence we see
that any fundamental relationship between $T_{c}$ and $\lambda^{-2}$
will just be a remapping of the superconducting dome. These results
are independent of the precise mechanism associated with the doping
process, which may in part be electrochemical, rather than purely
electrostatic. This suggests that ionic liquid gating may be a powerful
way to explore the phase diagram of other materials, and should be
considered to be an effective complement to conventional chemical
doping. 

The authors would like to thank Rafael Fernandes and Peter Orth for
useful discussions. This work was primarily supported by the National
Science Foundation through the University of Minnnesota MRSEC under
Award Number DMR-1420013 and partially supported by the National Science
Foundation under Award Number DMR-1209578. Part of this work was carried
out at the University of Minnesota Characterization Facility, a member
of the NSF-funded Materials Research Facilities Network via the MRSEC
program, and the Nanofabrication Center which receives partial support
from the NSF through the NNIN program.


\begin{thebibliography}{34}%
\makeatletter
\providecommand \@ifxundefined [1]{%
 \@ifx{#1\undefined}
}%
\providecommand \@ifnum [1]{%
 \ifnum #1\expandafter \@firstoftwo
 \else \expandafter \@secondoftwo
 \fi
}%
\providecommand \@ifx [1]{%
 \ifx #1\expandafter \@firstoftwo
 \else \expandafter \@secondoftwo
 \fi
}%
\providecommand \natexlab [1]{#1}%
\providecommand \enquote  [1]{``#1''}%
\providecommand \bibnamefont  [1]{#1}%
\providecommand \bibfnamefont [1]{#1}%
\providecommand \citenamefont [1]{#1}%
\providecommand \href@noop [0]{\@secondoftwo}%
\providecommand \href [0]{\begingroup \@sanitize@url \@href}%
\providecommand \@href[1]{\@@startlink{#1}\@@href}%
\providecommand \@@href[1]{\endgroup#1\@@endlink}%
\providecommand \@sanitize@url [0]{\catcode `\\12\catcode `\$12\catcode
  `\&12\catcode `\#12\catcode `\^12\catcode `\_12\catcode `\%12\relax}%
\providecommand \@@startlink[1]{}%
\providecommand \@@endlink[0]{}%
\providecommand \url  [0]{\begingroup\@sanitize@url \@url }%
\providecommand \@url [1]{\endgroup\@href {#1}{\urlprefix }}%
\providecommand \urlprefix  [0]{URL }%
\providecommand \Eprint [0]{\href }%
\providecommand \doibase [0]{http://dx.doi.org/}%
\providecommand \selectlanguage [0]{\@gobble}%
\providecommand \bibinfo  [0]{\@secondoftwo}%
\providecommand \bibfield  [0]{\@secondoftwo}%
\providecommand \translation [1]{[#1]}%
\providecommand \BibitemOpen [0]{}%
\providecommand \bibitemStop [0]{}%
\providecommand \bibitemNoStop [0]{.\EOS\space}%
\providecommand \EOS [0]{\spacefactor3000\relax}%
\providecommand \BibitemShut  [1]{\csname bibitem#1\endcsname}%
\let\auto@bib@innerbib\@empty
%</preamble>
\bibitem [{\citenamefont {Bednorz}\ and\ \citenamefont
  {M\"{u}ller}(1986)}]{Bednorz1986}%
  \BibitemOpen
  \bibfield  {author} {\bibinfo {author} {\bibfnamefont {J.~G.}\ \bibnamefont
  {Bednorz}}\ and\ \bibinfo {author} {\bibfnamefont {K.~A.}\ \bibnamefont
  {M\"{u}ller}},\ }\href {\doibase 10.1007/BF01303701} {\bibfield  {journal}
  {\bibinfo  {journal} {Zeitschrift f\"{u}r Physik B Condensed Matter}\
  }\textbf {\bibinfo {volume} {64}},\ \bibinfo {pages} {189} (\bibinfo {year}
  {1986})}\BibitemShut {NoStop}%
\bibitem [{\citenamefont {Homes}\ \emph {et~al.}(2004)\citenamefont {Homes},
  \citenamefont {Dordevic}, \citenamefont {Strongin}, \citenamefont {Bonn},
  \citenamefont {Liang}, \citenamefont {Hardy}, \citenamefont {Komiya},
  \citenamefont {Ando}, \citenamefont {Yu}, \citenamefont {Kaneko},
  \citenamefont {Zhao}, \citenamefont {Greven}, \citenamefont {Basov},\ and\
  \citenamefont {Timusk}}]{Homes2004}%
  \BibitemOpen
  \bibfield  {author} {\bibinfo {author} {\bibfnamefont {C.~C.}\ \bibnamefont
  {Homes}}, \bibinfo {author} {\bibfnamefont {S.~V.}\ \bibnamefont {Dordevic}},
  \bibinfo {author} {\bibfnamefont {M.}~\bibnamefont {Strongin}}, \bibinfo
  {author} {\bibfnamefont {D.~A.}\ \bibnamefont {Bonn}}, \bibinfo {author}
  {\bibfnamefont {R.}~\bibnamefont {Liang}}, \bibinfo {author} {\bibfnamefont
  {W.~N.}\ \bibnamefont {Hardy}}, \bibinfo {author} {\bibfnamefont
  {S.}~\bibnamefont {Komiya}}, \bibinfo {author} {\bibfnamefont
  {Y.}~\bibnamefont {Ando}}, \bibinfo {author} {\bibfnamefont {G.}~\bibnamefont
  {Yu}}, \bibinfo {author} {\bibfnamefont {N.}~\bibnamefont {Kaneko}}, \bibinfo
  {author} {\bibfnamefont {X.}~\bibnamefont {Zhao}}, \bibinfo {author}
  {\bibfnamefont {M.}~\bibnamefont {Greven}}, \bibinfo {author} {\bibfnamefont
  {D.~N.}\ \bibnamefont {Basov}}, \ and\ \bibinfo {author} {\bibfnamefont
  {T.}~\bibnamefont {Timusk}},\ }\href {\doibase 10.1038/nature02673}
  {\bibfield  {journal} {\bibinfo  {journal} {Nature}\ }\textbf {\bibinfo
  {volume} {430}},\ \bibinfo {pages} {539} (\bibinfo {year}
  {2004})}\BibitemShut {NoStop}%
\bibitem [{\citenamefont {Berkley}\ \emph {et~al.}(1988)\citenamefont
  {Berkley}, \citenamefont {Johnson}, \citenamefont {Anand}, \citenamefont
  {Beauchamp}, \citenamefont {Conroy}, \citenamefont {Goldman}, \citenamefont
  {Maps}, \citenamefont {Mauersberger}, \citenamefont {Mecartney},
  \citenamefont {Morton}, \citenamefont {Tuominen},\ and\ \citenamefont
  {Zhang}}]{Berkley1988}%
  \BibitemOpen
  \bibfield  {author} {\bibinfo {author} {\bibfnamefont {D.~D.}\ \bibnamefont
  {Berkley}}, \bibinfo {author} {\bibfnamefont {B.~R.}\ \bibnamefont
  {Johnson}}, \bibinfo {author} {\bibfnamefont {N.}~\bibnamefont {Anand}},
  \bibinfo {author} {\bibfnamefont {K.~M.}\ \bibnamefont {Beauchamp}}, \bibinfo
  {author} {\bibfnamefont {L.~E.}\ \bibnamefont {Conroy}}, \bibinfo {author}
  {\bibfnamefont {A.~M.}\ \bibnamefont {Goldman}}, \bibinfo {author}
  {\bibfnamefont {J.}~\bibnamefont {Maps}}, \bibinfo {author} {\bibfnamefont
  {K.}~\bibnamefont {Mauersberger}}, \bibinfo {author} {\bibfnamefont {M.~L.}\
  \bibnamefont {Mecartney}}, \bibinfo {author} {\bibfnamefont {J.}~\bibnamefont
  {Morton}}, \bibinfo {author} {\bibfnamefont {M.}~\bibnamefont {Tuominen}}, \
  and\ \bibinfo {author} {\bibfnamefont {Y.-J.}\ \bibnamefont {Zhang}},\ }\href
  {\doibase 10.1063/1.100489} {\bibfield  {journal} {\bibinfo  {journal}
  {Applied Physics Letters}\ }\textbf {\bibinfo {volume} {53}},\ \bibinfo
  {pages} {1973} (\bibinfo {year} {1988})}\BibitemShut {NoStop}%
\bibitem [{\citenamefont {Matthey}\ \emph {et~al.}(2007)\citenamefont
  {Matthey}, \citenamefont {Reyren}, \citenamefont {Triscone},\ and\
  \citenamefont {Schneider}}]{Matthey2007}%
  \BibitemOpen
  \bibfield  {author} {\bibinfo {author} {\bibfnamefont {D.}~\bibnamefont
  {Matthey}}, \bibinfo {author} {\bibfnamefont {N.}~\bibnamefont {Reyren}},
  \bibinfo {author} {\bibfnamefont {J.-M.~M.}\ \bibnamefont {Triscone}}, \ and\
  \bibinfo {author} {\bibfnamefont {T.}~\bibnamefont {Schneider}},\ }\href
  {\doibase 10.1103/PhysRevLett.98.057002} {\bibfield  {journal} {\bibinfo
  {journal} {Physical Review Letters}\ }\textbf {\bibinfo {volume} {98}},\
  \bibinfo {pages} {057002} (\bibinfo {year} {2007})}\BibitemShut {NoStop}%
\bibitem [{\citenamefont {Salluzzo}\ \emph {et~al.}(2008)\citenamefont
  {Salluzzo}, \citenamefont {Gambardella}, \citenamefont {{De Luca}},
  \citenamefont {{Di Capua}}, \citenamefont {Ristic},\ and\ \citenamefont
  {Vaglio}}]{Salluzzo2008}%
  \BibitemOpen
  \bibfield  {author} {\bibinfo {author} {\bibfnamefont {M.}~\bibnamefont
  {Salluzzo}}, \bibinfo {author} {\bibfnamefont {A.}~\bibnamefont
  {Gambardella}}, \bibinfo {author} {\bibfnamefont {G.~M.}\ \bibnamefont {{De
  Luca}}}, \bibinfo {author} {\bibfnamefont {R.}~\bibnamefont {{Di Capua}}},
  \bibinfo {author} {\bibfnamefont {Z.}~\bibnamefont {Ristic}}, \ and\ \bibinfo
  {author} {\bibfnamefont {R.}~\bibnamefont {Vaglio}},\ }\href {\doibase
  10.1103/PhysRevB.78.054524} {\bibfield  {journal} {\bibinfo  {journal}
  {Physical Review B}\ }\textbf {\bibinfo {volume} {78}},\ \bibinfo {pages}
  {054524} (\bibinfo {year} {2008})}\BibitemShut {NoStop}%
\bibitem [{\citenamefont {R\"{u}fenacht}\ \emph {et~al.}(2006)\citenamefont
  {R\"{u}fenacht}, \citenamefont {Locquet}, \citenamefont {Fompeyrine},
  \citenamefont {Caimi},\ and\ \citenamefont {Martinoli}}]{Rufenacht2006b}%
  \BibitemOpen
  \bibfield  {author} {\bibinfo {author} {\bibfnamefont {A.}~\bibnamefont
  {R\"{u}fenacht}}, \bibinfo {author} {\bibfnamefont {J.-P.~P.}\ \bibnamefont
  {Locquet}}, \bibinfo {author} {\bibfnamefont {J.}~\bibnamefont {Fompeyrine}},
  \bibinfo {author} {\bibfnamefont {D.}~\bibnamefont {Caimi}}, \ and\ \bibinfo
  {author} {\bibfnamefont {P.}~\bibnamefont {Martinoli}},\ }\href {\doibase
  10.1103/PhysRevLett.96.227002} {\bibfield  {journal} {\bibinfo  {journal}
  {Physical Review Letters}\ }\textbf {\bibinfo {volume} {96}},\ \bibinfo
  {pages} {227002} (\bibinfo {year} {2006})}\BibitemShut {NoStop}%
\bibitem [{\citenamefont {Ueno}\ \emph {et~al.}(2008)\citenamefont {Ueno},
  \citenamefont {Nakamura}, \citenamefont {Shimotani}, \citenamefont {Ohtomo},
  \citenamefont {Kimura}, \citenamefont {Nojima}, \citenamefont {Aoki},
  \citenamefont {Iwasa},\ and\ \citenamefont {Kawasaki}}]{Ueno2008}%
  \BibitemOpen
  \bibfield  {author} {\bibinfo {author} {\bibfnamefont {K.}~\bibnamefont
  {Ueno}}, \bibinfo {author} {\bibfnamefont {S.}~\bibnamefont {Nakamura}},
  \bibinfo {author} {\bibfnamefont {H.}~\bibnamefont {Shimotani}}, \bibinfo
  {author} {\bibfnamefont {A.}~\bibnamefont {Ohtomo}}, \bibinfo {author}
  {\bibfnamefont {N.}~\bibnamefont {Kimura}}, \bibinfo {author} {\bibfnamefont
  {T.}~\bibnamefont {Nojima}}, \bibinfo {author} {\bibfnamefont
  {H.}~\bibnamefont {Aoki}}, \bibinfo {author} {\bibfnamefont {Y.}~\bibnamefont
  {Iwasa}}, \ and\ \bibinfo {author} {\bibfnamefont {M.}~\bibnamefont
  {Kawasaki}},\ }\href {\doibase 10.1038/nmat2298} {\bibfield  {journal}
  {\bibinfo  {journal} {Nature materials}\ }\textbf {\bibinfo {volume} {7}},\
  \bibinfo {pages} {855} (\bibinfo {year} {2008})}\BibitemShut {NoStop}%
\bibitem [{\citenamefont {Bollinger}\ \emph {et~al.}(2011)\citenamefont
  {Bollinger}, \citenamefont {Dubuis}, \citenamefont {Yoon}, \citenamefont
  {Pavuna}, \citenamefont {Misewich},\ and\ \citenamefont
  {Bo\v{z}ovi\'{c}}}]{Bollinger2011}%
  \BibitemOpen
  \bibfield  {author} {\bibinfo {author} {\bibfnamefont {A.~T.}\ \bibnamefont
  {Bollinger}}, \bibinfo {author} {\bibfnamefont {G.}~\bibnamefont {Dubuis}},
  \bibinfo {author} {\bibfnamefont {J.}~\bibnamefont {Yoon}}, \bibinfo {author}
  {\bibfnamefont {D.}~\bibnamefont {Pavuna}}, \bibinfo {author} {\bibfnamefont
  {J.}~\bibnamefont {Misewich}}, \ and\ \bibinfo {author} {\bibfnamefont
  {I.}~\bibnamefont {Bo\v{z}ovi\'{c}}},\ }\href {\doibase 10.1038/nature09998}
  {\bibfield  {journal} {\bibinfo  {journal} {Nature}\ }\textbf {\bibinfo
  {volume} {472}},\ \bibinfo {pages} {458} (\bibinfo {year}
  {2011})}\BibitemShut {NoStop}%
\bibitem [{\citenamefont {Leng}\ \emph {et~al.}(2011)\citenamefont {Leng},
  \citenamefont {Garcia-Barriocanal}, \citenamefont {Bose}, \citenamefont
  {Lee},\ and\ \citenamefont {Goldman}}]{Leng2011}%
  \BibitemOpen
  \bibfield  {author} {\bibinfo {author} {\bibfnamefont {X.}~\bibnamefont
  {Leng}}, \bibinfo {author} {\bibfnamefont {J.}~\bibnamefont
  {Garcia-Barriocanal}}, \bibinfo {author} {\bibfnamefont {S.}~\bibnamefont
  {Bose}}, \bibinfo {author} {\bibfnamefont {Y.}~\bibnamefont {Lee}}, \ and\
  \bibinfo {author} {\bibfnamefont {A.~M.}\ \bibnamefont {Goldman}},\ }\href
  {\doibase 10.1103/PhysRevLett.107.027001} {\bibfield  {journal} {\bibinfo
  {journal} {Physical Review Letters}\ }\textbf {\bibinfo {volume} {107}},\
  \bibinfo {pages} {027001} (\bibinfo {year} {2011})}\BibitemShut {NoStop}%
\bibitem [{\citenamefont {Leng}\ \emph {et~al.}(2012)\citenamefont {Leng},
  \citenamefont {Garcia-Barriocanal}, \citenamefont {Yang}, \citenamefont
  {Lee}, \citenamefont {Kinney},\ and\ \citenamefont {Goldman}}]{Leng2012}%
  \BibitemOpen
  \bibfield  {author} {\bibinfo {author} {\bibfnamefont {X.}~\bibnamefont
  {Leng}}, \bibinfo {author} {\bibfnamefont {J.}~\bibnamefont
  {Garcia-Barriocanal}}, \bibinfo {author} {\bibfnamefont {B.}~\bibnamefont
  {Yang}}, \bibinfo {author} {\bibfnamefont {Y.}~\bibnamefont {Lee}}, \bibinfo
  {author} {\bibfnamefont {J.}~\bibnamefont {Kinney}}, \ and\ \bibinfo {author}
  {\bibfnamefont {A.~M.}\ \bibnamefont {Goldman}},\ }\href {\doibase
  10.1103/PhysRevLett.108.067004} {\bibfield  {journal} {\bibinfo  {journal}
  {Physical Review Letters}\ }\textbf {\bibinfo {volume} {108}},\ \bibinfo
  {pages} {067004} (\bibinfo {year} {2012})}\BibitemShut {NoStop}%
\bibitem [{\citenamefont {Garcia-Barriocanal}\ \emph
  {et~al.}(2013)\citenamefont {Garcia-Barriocanal}, \citenamefont {Kobrinskii},
  \citenamefont {Leng}, \citenamefont {Kinney}, \citenamefont {Yang},
  \citenamefont {Snyder},\ and\ \citenamefont
  {Goldman}}]{Garcia-Barriocanal2013}%
  \BibitemOpen
  \bibfield  {author} {\bibinfo {author} {\bibfnamefont {J.}~\bibnamefont
  {Garcia-Barriocanal}}, \bibinfo {author} {\bibfnamefont {A.}~\bibnamefont
  {Kobrinskii}}, \bibinfo {author} {\bibfnamefont {X.}~\bibnamefont {Leng}},
  \bibinfo {author} {\bibfnamefont {J.}~\bibnamefont {Kinney}}, \bibinfo
  {author} {\bibfnamefont {B.}~\bibnamefont {Yang}}, \bibinfo {author}
  {\bibfnamefont {S.}~\bibnamefont {Snyder}}, \ and\ \bibinfo {author}
  {\bibfnamefont {A.}~\bibnamefont {Goldman}},\ }\href {\doibase
  10.1103/PhysRevB.87.024509} {\bibfield  {journal} {\bibinfo  {journal}
  {Physical Review B}\ }\textbf {\bibinfo {volume} {87}},\ \bibinfo {pages}
  {024509} (\bibinfo {year} {2013})}\BibitemShut {NoStop}%
\bibitem [{\citenamefont {Zeng}\ \emph {et~al.}(2015)\citenamefont {Zeng},
  \citenamefont {Huang}, \citenamefont {Lv}, \citenamefont {Bao}, \citenamefont
  {Gopinadhan}, \citenamefont {Jian}, \citenamefont {Herng}, \citenamefont
  {Liu}, \citenamefont {Zhao}, \citenamefont {Li}, \citenamefont {{Harsan Ma}},
  \citenamefont {Yang}, \citenamefont {Ding}, \citenamefont {Venkatesan},\ and\
  \citenamefont {Ariando}}]{Zeng2015}%
  \BibitemOpen
  \bibfield  {author} {\bibinfo {author} {\bibfnamefont {S.~W.}\ \bibnamefont
  {Zeng}}, \bibinfo {author} {\bibfnamefont {Z.}~\bibnamefont {Huang}},
  \bibinfo {author} {\bibfnamefont {W.~M.}\ \bibnamefont {Lv}}, \bibinfo
  {author} {\bibfnamefont {N.~N.}\ \bibnamefont {Bao}}, \bibinfo {author}
  {\bibfnamefont {K.}~\bibnamefont {Gopinadhan}}, \bibinfo {author}
  {\bibfnamefont {L.~K.}\ \bibnamefont {Jian}}, \bibinfo {author}
  {\bibfnamefont {T.~S.}\ \bibnamefont {Herng}}, \bibinfo {author}
  {\bibfnamefont {Z.~Q.}\ \bibnamefont {Liu}}, \bibinfo {author} {\bibfnamefont
  {Y.~L.}\ \bibnamefont {Zhao}}, \bibinfo {author} {\bibfnamefont {C.~J.}\
  \bibnamefont {Li}}, \bibinfo {author} {\bibfnamefont {H.~J.}\ \bibnamefont
  {{Harsan Ma}}}, \bibinfo {author} {\bibfnamefont {P.}~\bibnamefont {Yang}},
  \bibinfo {author} {\bibfnamefont {J.}~\bibnamefont {Ding}}, \bibinfo {author}
  {\bibfnamefont {T.}~\bibnamefont {Venkatesan}}, \ and\ \bibinfo {author}
  {\bibnamefont {Ariando}},\ }\href {\doibase 10.1103/PhysRevB.92.020503}
  {\bibfield  {journal} {\bibinfo  {journal} {Physical Review B}\ }\textbf
  {\bibinfo {volume} {92}},\ \bibinfo {pages} {020503} (\bibinfo {year}
  {2015})}\BibitemShut {NoStop}%
\bibitem [{\citenamefont {Hebard}\ and\ \citenamefont
  {Fiory}(1980)}]{Hebard1980}%
  \BibitemOpen
  \bibfield  {author} {\bibinfo {author} {\bibfnamefont {A.}~\bibnamefont
  {Hebard}}\ and\ \bibinfo {author} {\bibfnamefont {A.}~\bibnamefont {Fiory}},\
  }\href {\doibase 10.1103/PhysRevLett.44.291} {\bibfield  {journal} {\bibinfo
  {journal} {Physical Review Letters}\ }\textbf {\bibinfo {volume} {44}},\
  \bibinfo {pages} {291} (\bibinfo {year} {1980})}\BibitemShut {NoStop}%
\bibitem [{\citenamefont {Jeanneret}\ \emph {et~al.}(1989)\citenamefont
  {Jeanneret}, \citenamefont {Gavilano}, \citenamefont {Racine}, \citenamefont
  {Leemann},\ and\ \citenamefont {Martinoli}}]{Jeanneret1989}%
  \BibitemOpen
  \bibfield  {author} {\bibinfo {author} {\bibfnamefont {B.}~\bibnamefont
  {Jeanneret}}, \bibinfo {author} {\bibfnamefont {J.~L.}\ \bibnamefont
  {Gavilano}}, \bibinfo {author} {\bibfnamefont {G.~A.}\ \bibnamefont
  {Racine}}, \bibinfo {author} {\bibfnamefont {C.}~\bibnamefont {Leemann}}, \
  and\ \bibinfo {author} {\bibfnamefont {P.}~\bibnamefont {Martinoli}},\ }\href
  {\doibase 10.1063/1.102053} {\bibfield  {journal} {\bibinfo  {journal}
  {Applied Physics Letters}\ }\textbf {\bibinfo {volume} {55}},\ \bibinfo
  {pages} {2336} (\bibinfo {year} {1989})}\BibitemShut {NoStop}%
\bibitem [{\citenamefont {Lin}\ \emph {et~al.}(1995)\citenamefont {Lin},
  \citenamefont {Spalding}, \citenamefont {Goldman}, \citenamefont {Bayman},\
  and\ \citenamefont {Valls}}]{Lin1995a}%
  \BibitemOpen
  \bibfield  {author} {\bibinfo {author} {\bibfnamefont {Z.-H.}\ \bibnamefont
  {Lin}}, \bibinfo {author} {\bibfnamefont {G.~C.}\ \bibnamefont {Spalding}},
  \bibinfo {author} {\bibfnamefont {A.~M.}\ \bibnamefont {Goldman}}, \bibinfo
  {author} {\bibfnamefont {B.~F.}\ \bibnamefont {Bayman}}, \ and\ \bibinfo
  {author} {\bibfnamefont {O.~T.}\ \bibnamefont {Valls}},\ }\href {\doibase
  10.1209/0295-5075/32/7/006} {\bibfield  {journal} {\bibinfo  {journal}
  {Europhysics Letters (EPL)}\ }\textbf {\bibinfo {volume} {32}},\ \bibinfo
  {pages} {573} (\bibinfo {year} {1995})}\BibitemShut {NoStop}%
\bibitem [{\citenamefont {Turneaure}\ \emph {et~al.}(1996)\citenamefont
  {Turneaure}, \citenamefont {Ulm},\ and\ \citenamefont
  {Lemberger}}]{Turneaure1996}%
  \BibitemOpen
  \bibfield  {author} {\bibinfo {author} {\bibfnamefont {S.~J.}\ \bibnamefont
  {Turneaure}}, \bibinfo {author} {\bibfnamefont {E.~R.}\ \bibnamefont {Ulm}},
  \ and\ \bibinfo {author} {\bibfnamefont {T.~R.}\ \bibnamefont {Lemberger}},\
  }\href {\doibase 10.1063/1.362657} {\bibfield  {journal} {\bibinfo  {journal}
  {Journal of Applied Physics}\ }\textbf {\bibinfo {volume} {79}},\ \bibinfo
  {pages} {4221} (\bibinfo {year} {1996})}\BibitemShut {NoStop}%
\bibitem [{\citenamefont {Turneaure}\ \emph {et~al.}(1998)\citenamefont
  {Turneaure}, \citenamefont {Pesetski},\ and\ \citenamefont
  {Lemberger}}]{Turneaure1998}%
  \BibitemOpen
  \bibfield  {author} {\bibinfo {author} {\bibfnamefont {S.~J.}\ \bibnamefont
  {Turneaure}}, \bibinfo {author} {\bibfnamefont {A.~a.}\ \bibnamefont
  {Pesetski}}, \ and\ \bibinfo {author} {\bibfnamefont {T.~R.}\ \bibnamefont
  {Lemberger}},\ }\href {\doibase 10.1063/1.367193} {\bibfield  {journal}
  {\bibinfo  {journal} {Journal of Applied Physics}\ }\textbf {\bibinfo
  {volume} {83}},\ \bibinfo {pages} {4334} (\bibinfo {year}
  {1998})}\BibitemShut {NoStop}%
\bibitem [{\citenamefont {Jeong}\ \emph {et~al.}(2013)\citenamefont {Jeong},
  \citenamefont {Aetukuri}, \citenamefont {Graf}, \citenamefont {Schladt},
  \citenamefont {Samant},\ and\ \citenamefont {Parkin}}]{Jeong2013}%
  \BibitemOpen
  \bibfield  {author} {\bibinfo {author} {\bibfnamefont {J.}~\bibnamefont
  {Jeong}}, \bibinfo {author} {\bibfnamefont {N.}~\bibnamefont {Aetukuri}},
  \bibinfo {author} {\bibfnamefont {T.}~\bibnamefont {Graf}}, \bibinfo {author}
  {\bibfnamefont {T.~D.}\ \bibnamefont {Schladt}}, \bibinfo {author}
  {\bibfnamefont {M.~G.}\ \bibnamefont {Samant}}, \ and\ \bibinfo {author}
  {\bibfnamefont {S.~S.~P.}\ \bibnamefont {Parkin}},\ }\href {\doibase
  10.1126/science.1230512} {\bibfield  {journal} {\bibinfo  {journal} {Science
  (New York, N.Y.)}\ }\textbf {\bibinfo {volume} {339}},\ \bibinfo {pages}
  {1402} (\bibinfo {year} {2013})}\BibitemShut {NoStop}%
\bibitem [{\citenamefont {Jeong}\ \emph {et~al.}(2015)\citenamefont {Jeong},
  \citenamefont {Aetukuri}, \citenamefont {Passarello}, \citenamefont
  {Conradson}, \citenamefont {Samant},\ and\ \citenamefont
  {Parkin}}]{Jeong2015}%
  \BibitemOpen
  \bibfield  {author} {\bibinfo {author} {\bibfnamefont {J.}~\bibnamefont
  {Jeong}}, \bibinfo {author} {\bibfnamefont {N.~B.}\ \bibnamefont {Aetukuri}},
  \bibinfo {author} {\bibfnamefont {D.}~\bibnamefont {Passarello}}, \bibinfo
  {author} {\bibfnamefont {S.~D.}\ \bibnamefont {Conradson}}, \bibinfo {author}
  {\bibfnamefont {M.~G.}\ \bibnamefont {Samant}}, \ and\ \bibinfo {author}
  {\bibfnamefont {S.~S.~P.}\ \bibnamefont {Parkin}},\ }\href {\doibase
  10.1073/pnas.1419051112} {\bibfield  {journal} {\bibinfo  {journal}
  {Proceedings of the National Academy of Sciences}\ }\textbf {\bibinfo
  {volume} {112}},\ \bibinfo {pages} {1013} (\bibinfo {year}
  {2015})}\BibitemShut {NoStop}%
\bibitem [{\citenamefont {Lang}\ \emph {et~al.}(2014)\citenamefont {Lang},
  \citenamefont {Sloppy}, \citenamefont {Ghassemi}, \citenamefont {Devlin},
  \citenamefont {Sichel-Tissot}, \citenamefont {Idrobo}, \citenamefont {May},\
  and\ \citenamefont {Taheri}}]{Lang2014}%
  \BibitemOpen
  \bibfield  {author} {\bibinfo {author} {\bibfnamefont {A.~C.}\ \bibnamefont
  {Lang}}, \bibinfo {author} {\bibfnamefont {J.~D.}\ \bibnamefont {Sloppy}},
  \bibinfo {author} {\bibfnamefont {H.}~\bibnamefont {Ghassemi}}, \bibinfo
  {author} {\bibfnamefont {R.~C.}\ \bibnamefont {Devlin}}, \bibinfo {author}
  {\bibfnamefont {R.~J.}\ \bibnamefont {Sichel-Tissot}}, \bibinfo {author}
  {\bibfnamefont {J.-C.}\ \bibnamefont {Idrobo}}, \bibinfo {author}
  {\bibfnamefont {S.~J.}\ \bibnamefont {May}}, \ and\ \bibinfo {author}
  {\bibfnamefont {M.~L.}\ \bibnamefont {Taheri}},\ }\href {\doibase
  10.1021/am504547b} {\bibfield  {journal} {\bibinfo  {journal} {ACS Applied
  Materials \& Interfaces}\ }\textbf {\bibinfo {volume} {6}},\ \bibinfo {pages}
  {17018} (\bibinfo {year} {2014})}\BibitemShut {NoStop}%
\bibitem [{\citenamefont {Petach}\ \emph {et~al.}(2014)\citenamefont {Petach},
  \citenamefont {Lee}, \citenamefont {Davis}, \citenamefont {Mehta},\ and\
  \citenamefont {Goldhaber-Gordon}}]{Petach2014}%
  \BibitemOpen
  \bibfield  {author} {\bibinfo {author} {\bibfnamefont {T.~A.}\ \bibnamefont
  {Petach}}, \bibinfo {author} {\bibfnamefont {M.}~\bibnamefont {Lee}},
  \bibinfo {author} {\bibfnamefont {R.~C.}\ \bibnamefont {Davis}}, \bibinfo
  {author} {\bibfnamefont {A.}~\bibnamefont {Mehta}}, \ and\ \bibinfo {author}
  {\bibfnamefont {D.}~\bibnamefont {Goldhaber-Gordon}},\ }\href {\doibase
  10.1103/PhysRevB.90.081108} {\bibfield  {journal} {\bibinfo  {journal}
  {Physical Review B}\ }\textbf {\bibinfo {volume} {90}},\ \bibinfo {pages}
  {081108(R)} (\bibinfo {year} {2014})}\BibitemShut {NoStop}%
\bibitem [{\citenamefont {Fratini}\ \emph {et~al.}(2010)\citenamefont
  {Fratini}, \citenamefont {Poccia}, \citenamefont {Ricci}, \citenamefont
  {Campi}, \citenamefont {Burghammer}, \citenamefont {Aeppli},\ and\
  \citenamefont {Bianconi}}]{Fratini2010}%
  \BibitemOpen
  \bibfield  {author} {\bibinfo {author} {\bibfnamefont {M.}~\bibnamefont
  {Fratini}}, \bibinfo {author} {\bibfnamefont {N.}~\bibnamefont {Poccia}},
  \bibinfo {author} {\bibfnamefont {A.}~\bibnamefont {Ricci}}, \bibinfo
  {author} {\bibfnamefont {G.}~\bibnamefont {Campi}}, \bibinfo {author}
  {\bibfnamefont {M.}~\bibnamefont {Burghammer}}, \bibinfo {author}
  {\bibfnamefont {G.}~\bibnamefont {Aeppli}}, \ and\ \bibinfo {author}
  {\bibfnamefont {A.}~\bibnamefont {Bianconi}},\ }\href {\doibase
  10.1038/nature09260} {\bibfield  {journal} {\bibinfo  {journal} {Nature}\
  }\textbf {\bibinfo {volume} {466}},\ \bibinfo {pages} {841} (\bibinfo {year}
  {2010})}\BibitemShut {NoStop}%
\bibitem [{\citenamefont {Startseva}\ \emph {et~al.}(1998)\citenamefont
  {Startseva}, \citenamefont {Timusk}, \citenamefont {Puchkov}, \citenamefont
  {Basov}, \citenamefont {Mook}, \citenamefont {Okuya}, \citenamefont
  {Kimura},\ and\ \citenamefont {Kishio}}]{Startseva1998}%
  \BibitemOpen
  \bibfield  {author} {\bibinfo {author} {\bibfnamefont {T.}~\bibnamefont
  {Startseva}}, \bibinfo {author} {\bibfnamefont {T.}~\bibnamefont {Timusk}},
  \bibinfo {author} {\bibfnamefont {A.~V.}\ \bibnamefont {Puchkov}}, \bibinfo
  {author} {\bibfnamefont {D.~N.}\ \bibnamefont {Basov}}, \bibinfo {author}
  {\bibfnamefont {H.~A.}\ \bibnamefont {Mook}}, \bibinfo {author}
  {\bibfnamefont {M.}~\bibnamefont {Okuya}}, \bibinfo {author} {\bibfnamefont
  {T.}~\bibnamefont {Kimura}}, \ and\ \bibinfo {author} {\bibfnamefont
  {K.}~\bibnamefont {Kishio}},\ }\href {\doibase 10.1103/PhysRevB.59.7184}
  {\bibfield  {journal} {\bibinfo  {journal} {Physical Review B}\ }\textbf
  {\bibinfo {volume} {59}},\ \bibinfo {pages} {8} (\bibinfo {year}
  {1998})}\BibitemShut {NoStop}%
\bibitem [{\citenamefont {Aeppli}\ \emph {et~al.}(1987)\citenamefont {Aeppli},
  \citenamefont {Ansaldo}, \citenamefont {Brewer}, \citenamefont {Cava},
  \citenamefont {Kiefl}, \citenamefont {Kreitzman}, \citenamefont {Luke},\ and\
  \citenamefont {Noakes}}]{Aeppli1987}%
  \BibitemOpen
  \bibfield  {author} {\bibinfo {author} {\bibfnamefont {G.}~\bibnamefont
  {Aeppli}}, \bibinfo {author} {\bibfnamefont {E.~J.}\ \bibnamefont {Ansaldo}},
  \bibinfo {author} {\bibfnamefont {J.~H.}\ \bibnamefont {Brewer}}, \bibinfo
  {author} {\bibfnamefont {R.~J.}\ \bibnamefont {Cava}}, \bibinfo {author}
  {\bibfnamefont {R.~F.}\ \bibnamefont {Kiefl}}, \bibinfo {author}
  {\bibfnamefont {S.~R.}\ \bibnamefont {Kreitzman}}, \bibinfo {author}
  {\bibfnamefont {G.~M.}\ \bibnamefont {Luke}}, \ and\ \bibinfo {author}
  {\bibfnamefont {D.~R.}\ \bibnamefont {Noakes}},\ }\href {\doibase
  10.1103/PhysRevB.35.7129} {\bibfield  {journal} {\bibinfo  {journal}
  {Physical Review B}\ }\textbf {\bibinfo {volume} {35}},\ \bibinfo {pages}
  {7129} (\bibinfo {year} {1987})}\BibitemShut {NoStop}%
\bibitem [{\citenamefont {Lemberger}\ \emph {et~al.}(2011)\citenamefont
  {Lemberger}, \citenamefont {Hetel}, \citenamefont {Tsukada}, \citenamefont
  {Naito},\ and\ \citenamefont {Randeria}}]{Lemberger2011}%
  \BibitemOpen
  \bibfield  {author} {\bibinfo {author} {\bibfnamefont {T.~R.}\ \bibnamefont
  {Lemberger}}, \bibinfo {author} {\bibfnamefont {I.}~\bibnamefont {Hetel}},
  \bibinfo {author} {\bibfnamefont {A.}~\bibnamefont {Tsukada}}, \bibinfo
  {author} {\bibfnamefont {M.}~\bibnamefont {Naito}}, \ and\ \bibinfo {author}
  {\bibfnamefont {M.}~\bibnamefont {Randeria}},\ }\href {\doibase
  10.1103/PhysRevB.83.140507} {\bibfield  {journal} {\bibinfo  {journal}
  {Physical Review B}\ }\textbf {\bibinfo {volume} {83}},\ \bibinfo {pages}
  {140507} (\bibinfo {year} {2011})}\BibitemShut {NoStop}%
\bibitem [{\citenamefont {Prohammer}\ and\ \citenamefont
  {Carbotte}(1991)}]{Prohammer1991}%
  \BibitemOpen
  \bibfield  {author} {\bibinfo {author} {\bibfnamefont {M.}~\bibnamefont
  {Prohammer}}\ and\ \bibinfo {author} {\bibfnamefont {J.~P.}\ \bibnamefont
  {Carbotte}},\ }\href {\doibase 10.1103/PhysRevB.43.5370} {\bibfield
  {journal} {\bibinfo  {journal} {Physical Review B}\ }\textbf {\bibinfo
  {volume} {43}},\ \bibinfo {pages} {5370} (\bibinfo {year}
  {1991})}\BibitemShut {NoStop}%
\bibitem [{\citenamefont {Homes}\ \emph {et~al.}(2005)\citenamefont {Homes},
  \citenamefont {Dordevic}, \citenamefont {Valla},\ and\ \citenamefont
  {Strongin}}]{Homes2005}%
  \BibitemOpen
  \bibfield  {author} {\bibinfo {author} {\bibfnamefont {C.~C.}\ \bibnamefont
  {Homes}}, \bibinfo {author} {\bibfnamefont {S.~V.}\ \bibnamefont {Dordevic}},
  \bibinfo {author} {\bibfnamefont {T.}~\bibnamefont {Valla}}, \ and\ \bibinfo
  {author} {\bibfnamefont {M.}~\bibnamefont {Strongin}},\ }\href {\doibase
  10.1103/PhysRevB.72.134517} {\bibfield  {journal} {\bibinfo  {journal}
  {Physical Review B}\ }\textbf {\bibinfo {volume} {72}},\ \bibinfo {pages}
  {134517} (\bibinfo {year} {2005})}\BibitemShut {NoStop}%
\bibitem [{\citenamefont {Zuev}\ \emph {et~al.}(2005)\citenamefont {Zuev},
  \citenamefont {{Seog Kim}},\ and\ \citenamefont {Lemberger}}]{Zuev2005}%
  \BibitemOpen
  \bibfield  {author} {\bibinfo {author} {\bibfnamefont {Y.}~\bibnamefont
  {Zuev}}, \bibinfo {author} {\bibfnamefont {M.}~\bibnamefont {{Seog Kim}}}, \
  and\ \bibinfo {author} {\bibfnamefont {T.~R.}\ \bibnamefont {Lemberger}},\
  }\href {\doibase 10.1103/PhysRevLett.95.137002} {\bibfield  {journal}
  {\bibinfo  {journal} {Physical Review Letters}\ }\textbf {\bibinfo {volume}
  {95}},\ \bibinfo {pages} {137002} (\bibinfo {year} {2005})}\BibitemShut
  {NoStop}%
\bibitem [{\citenamefont {Imry}\ \emph {et~al.}(2012)\citenamefont {Imry},
  \citenamefont {Strongin},\ and\ \citenamefont {Homes}}]{Imry2012}%
  \BibitemOpen
  \bibfield  {author} {\bibinfo {author} {\bibfnamefont {Y.}~\bibnamefont
  {Imry}}, \bibinfo {author} {\bibfnamefont {M.}~\bibnamefont {Strongin}}, \
  and\ \bibinfo {author} {\bibfnamefont {C.~C.}\ \bibnamefont {Homes}},\ }\href
  {\doibase 10.1103/PhysRevLett.109.067003} {\bibfield  {journal} {\bibinfo
  {journal} {Physical Review Letters}\ }\textbf {\bibinfo {volume} {109}},\
  \bibinfo {pages} {067003} (\bibinfo {year} {2012})}\BibitemShut {NoStop}%
\bibitem [{\citenamefont {Kogan}(2013)}]{Kogan2013}%
  \BibitemOpen
  \bibfield  {author} {\bibinfo {author} {\bibfnamefont {V.~G.}\ \bibnamefont
  {Kogan}},\ }\href {\doibase 10.1103/PhysRevB.87.220507} {\bibfield  {journal}
  {\bibinfo  {journal} {Physical Review B}\ }\textbf {\bibinfo {volume} {87}},\
  \bibinfo {pages} {220507} (\bibinfo {year} {2013})}\BibitemShut {NoStop}%
\bibitem [{\citenamefont {Taylor}\ and\ \citenamefont
  {Maple}(2007)}]{Taylor2007}%
  \BibitemOpen
  \bibfield  {author} {\bibinfo {author} {\bibfnamefont {B.~J.}\ \bibnamefont
  {Taylor}}\ and\ \bibinfo {author} {\bibfnamefont {M.~B.}\ \bibnamefont
  {Maple}},\ }\href {\doibase 10.1103/PhysRevB.76.184512} {\bibfield  {journal}
  {\bibinfo  {journal} {Physical Review B}\ }\textbf {\bibinfo {volume} {76}},\
  \bibinfo {pages} {184512} (\bibinfo {year} {2007})}\BibitemShut {NoStop}%
\bibitem [{\citenamefont {Basov}\ and\ \citenamefont
  {Chubukov}(2011)}]{Basov2011}%
  \BibitemOpen
  \bibfield  {author} {\bibinfo {author} {\bibfnamefont {D.~N.}\ \bibnamefont
  {Basov}}\ and\ \bibinfo {author} {\bibfnamefont {A.~V.}\ \bibnamefont
  {Chubukov}},\ }\href {\doibase 10.1038/nphys1975} {\bibfield  {journal}
  {\bibinfo  {journal} {Nature Physics}\ }\textbf {\bibinfo {volume} {7}},\
  \bibinfo {pages} {272} (\bibinfo {year} {2011})}\BibitemShut {NoStop}%
\bibitem [{\citenamefont {Hashimoto}\ \emph {et~al.}(2014)\citenamefont
  {Hashimoto}, \citenamefont {Vishik}, \citenamefont {He}, \citenamefont
  {Devereaux},\ and\ \citenamefont {Shen}}]{Hashimoto2014}%
  \BibitemOpen
  \bibfield  {author} {\bibinfo {author} {\bibfnamefont {M.}~\bibnamefont
  {Hashimoto}}, \bibinfo {author} {\bibfnamefont {I.~M.}\ \bibnamefont
  {Vishik}}, \bibinfo {author} {\bibfnamefont {R.-H.}\ \bibnamefont {He}},
  \bibinfo {author} {\bibfnamefont {T.~P.}\ \bibnamefont {Devereaux}}, \ and\
  \bibinfo {author} {\bibfnamefont {Z.-X.}\ \bibnamefont {Shen}},\ }\href
  {\doibase 10.1038/nphys3009} {\bibfield  {journal} {\bibinfo  {journal}
  {Nature Physics}\ }\textbf {\bibinfo {volume} {10}},\ \bibinfo {pages} {483}
  (\bibinfo {year} {2014})}\BibitemShut {NoStop}%
\bibitem [{\citenamefont {Uemura}\ \emph {et~al.}(1989)\citenamefont {Uemura},
  \citenamefont {Luke}, \citenamefont {Sternlieb}, \citenamefont {Brewer},
  \citenamefont {Carolan}, \citenamefont {Hardy}, \citenamefont {Kadono},
  \citenamefont {Kempton}, \citenamefont {Kiefl}, \citenamefont {Kreitzman},
  \citenamefont {Mulhern}, \citenamefont {Riseman}, \citenamefont {Williams},
  \citenamefont {Yang}, \citenamefont {Uchida}, \citenamefont {Takagi},
  \citenamefont {Gopalakrishnan}, \citenamefont {Sleight}, \citenamefont
  {Subramanian}, \citenamefont {Chien}, \citenamefont {Cieplak}, \citenamefont
  {Xiao}, \citenamefont {Lee}, \citenamefont {Statt}, \citenamefont {Stronach},
  \citenamefont {Kossler},\ and\ \citenamefont {Yu}}]{Uemura1989}%
  \BibitemOpen
  \bibfield  {author} {\bibinfo {author} {\bibfnamefont {Y.}~\bibnamefont
  {Uemura}}, \bibinfo {author} {\bibfnamefont {G.}~\bibnamefont {Luke}},
  \bibinfo {author} {\bibfnamefont {B.}~\bibnamefont {Sternlieb}}, \bibinfo
  {author} {\bibfnamefont {J.}~\bibnamefont {Brewer}}, \bibinfo {author}
  {\bibfnamefont {J.}~\bibnamefont {Carolan}}, \bibinfo {author} {\bibfnamefont
  {W.}~\bibnamefont {Hardy}}, \bibinfo {author} {\bibfnamefont
  {R.}~\bibnamefont {Kadono}}, \bibinfo {author} {\bibfnamefont
  {J.}~\bibnamefont {Kempton}}, \bibinfo {author} {\bibfnamefont
  {R.}~\bibnamefont {Kiefl}}, \bibinfo {author} {\bibfnamefont
  {S.}~\bibnamefont {Kreitzman}}, \bibinfo {author} {\bibfnamefont
  {P.}~\bibnamefont {Mulhern}}, \bibinfo {author} {\bibfnamefont
  {T.}~\bibnamefont {Riseman}}, \bibinfo {author} {\bibfnamefont
  {D.}~\bibnamefont {Williams}}, \bibinfo {author} {\bibfnamefont
  {B.}~\bibnamefont {Yang}}, \bibinfo {author} {\bibfnamefont {S.}~\bibnamefont
  {Uchida}}, \bibinfo {author} {\bibfnamefont {H.}~\bibnamefont {Takagi}},
  \bibinfo {author} {\bibfnamefont {J.}~\bibnamefont {Gopalakrishnan}},
  \bibinfo {author} {\bibfnamefont {A.}~\bibnamefont {Sleight}}, \bibinfo
  {author} {\bibfnamefont {M.}~\bibnamefont {Subramanian}}, \bibinfo {author}
  {\bibfnamefont {C.}~\bibnamefont {Chien}}, \bibinfo {author} {\bibfnamefont
  {M.}~\bibnamefont {Cieplak}}, \bibinfo {author} {\bibfnamefont
  {G.}~\bibnamefont {Xiao}}, \bibinfo {author} {\bibfnamefont {V.}~\bibnamefont
  {Lee}}, \bibinfo {author} {\bibfnamefont {B.}~\bibnamefont {Statt}}, \bibinfo
  {author} {\bibfnamefont {C.}~\bibnamefont {Stronach}}, \bibinfo {author}
  {\bibfnamefont {W.}~\bibnamefont {Kossler}}, \ and\ \bibinfo {author}
  {\bibfnamefont {X.}~\bibnamefont {Yu}},\ }\href {\doibase
  10.1103/PhysRevLett.62.2317} {\bibfield  {journal} {\bibinfo  {journal}
  {Physical Review Letters}\ }\textbf {\bibinfo {volume} {62}},\ \bibinfo
  {pages} {2317} (\bibinfo {year} {1989})}\BibitemShut {NoStop}%
\end{thebibliography}
\end{document}